\documentclass[smallextended]{svjour3} 
\usepackage[english]{babel}
\usepackage[utf8]{inputenc}
\usepackage{pgf,tikz,pgfplots}
\pgfplotsset{compat=1.15}
\usetikzlibrary{arrows}
\usepackage{amssymb}
\usepackage{bm}
\usepackage{csquotes}
\expandafter\let\csname equation*\endcsname\relax
\expandafter\let\csname endequation*\endcsname\relax
\usepackage{amsmath}
\usepackage{amsfonts}
\usepackage{amstext}
\usepackage{amssymb}
\usepackage{physics}
\usepackage{nicematrix}
\usepackage{amsmath}
\usepackage{amsfonts}
\usepackage{amssymb}

\usepackage{physics}
\usepackage{color}
\usepackage{hyperref}
\usepackage{enumerate}
\usepackage{graphicx}
\usepackage{float}
\usepackage{amsmath}
\usepackage{amsfonts}
\usepackage{setspace} 
\usepackage{lipsum}
\usepackage{dcolumn}
\usepackage{hyperref}
\usepackage{subfigure}
\usepackage{epsfig}
\usepackage{epsf}
\usepackage{epstopdf}

\usepackage{color}
\usepackage{enumitem}
\usepackage{bm}
\usepackage{subfigure}
\usepackage{xcolor}
\usepackage{tikz}
\usepackage[margin=1in]{geometry}
\usepackage{pgfplots}
\definecolor{lightgray}{gray}{.90}
\definecolor{darkred}{rgb}{0.9,0.1,0.1}

\newcommand\correspondingauthor{\thanks{Corresponding author: C. Bernardin (cbernard@unice.fr)}}

\begin{document}

\title{Heat transport in an ordered  harmonic chain in presence of a uniform magnetic field}

\author{Junaid Majeed Bhat$^1$, Gaëtan Cane$^2$, Cédric Bernardin$^{2,3}$\correspondingauthor and Abhishek Dhar$^1$}

\institute{${}^{1}$ International Centre for Theoretical Sciences, Bengaluru, India-560089 \\
	${}^{2}$ Universit\'e C\^ote d'Azur, CNRS, LJAD, Parc Valrose, 06108 Nice Cedex 02, France\\
	${}^{3}$  Interdisciplinary Scientific Center Poncelet (CNRS IRL 2615), 119002 Moscow, Russia}
\date{Received: date / Accepted: date}

\maketitle

\begin{abstract}
We consider heat transport across a  harmonic chain of charged particles, with transverse degrees of freedom,  in the presence of a uniform magnetic field. For an open chain connected to  heat baths at the two ends we obtain the nonequilibrium Green's function  expression for the heat current. This expression involves two different Green's functions which can be identified as corresponding respectively to scattering processes within or between the two transverse waves.  The presence of the magnetic field leads to two phonon bands of the isolated system and  we show that the net transmission can be written as a sum of two distinct terms attributable to the two bands. Exact expressions are obtained for the current  in the thermodynamic limit, for the the cases of free and fixed boundary conditions. In this limit, we find that  at small frequency $\omega$,  the effective transmission has the frequency-dependence $\omega^{3/2}$ and $\omega^{1/2}$ for fixed and free boundary conditions respectively. This is in contrast to the zero magnetic field case where  the transmission has the dependence $\omega^2$ and $\omega^0$ for the two boundary conditions respectively, and can be understood as arising from the quadratic low frequency phonon dispersion.    
\end{abstract}

%
%
%
%
%
\tableofcontents

\section{Introduction}
Heat transport in  harmonic chains connected to heat reservoirs has been extensively studied since the seminal work of Rieder, Lebowitz and Lieb~\cite{rieder1967properties} (RLL).  The main results of RLL were the demonstration  that, in the nonequilibrium steady state (NESS), the heat current  across the chain saturates with increase in system size while the temperature profile is flat in the bulk of the chain.  They provided  exact expressions for the asymptotic values of the current and the temperature profile. In later work Nakazawa extended these results to systems with on-site potentials, not considered by RLL, and to higher dimensions~\cite{nakazawa1968energy,nakazawa1970lattice}. RLL and Nakazawa obtained the steady state current by an exact solution of the correlation matrix which described the Gaussian NESS measure. An alternative approach to compute NESS properties of harmonic chains was used by~\cite{casher1971heat,roy2008heat} and this is essentially based on the so called  Landauer and nonequilibrium Green's function formalism (NEGF) where the current is expressed as an integral over frequency of a phonon transmission coefficient.  The NEGF approach has been sucessfully applied to various classical and quantum systems~\cite{dhar2001heat,dhar2006heat,roy2008heat,kannan2012,dhar2006nonequilibrium,bhat2020transport}.

Recently,  interesting results were obtained for heat transfer in harmonic chains made of charged atoms that interact with an external magnetic field~\cite{tamaki2017heat,saito2018,tamaki2018nernst}.  In particular, Ref.~\cite{tamaki2017heat} studied a harmonic chain with transverse degrees of freedom and with the Hamiltonian dynamics perturbed by stochastic noise that conserves both momentum and energy. This perturbation is added to micmic the deterministic chaos which should be produced by the nonlinearities of the interactions if they were present \cite{BBJKO}. In the context of anomalous heat transport the authors identify a new universality class based on the system size dependence of the thermal conductivity. 
 An interesting observation made in the paper is on the phonon dispersion in this model with a finite magnetic field --- out of the two phonon bands it was found that the lower band has a low frequency dispersion $\omega \sim q^2$ and so a vanishing sound speed. A natural question is the effect of this on heat conduction in this system in the absence of stochastic noise. This question is addressed in the present paper. Systems of quantum harmonic oscillator chains  in external  magnetic fields and in thermal equilibrium have earlier been studied in the context of ergodicity, in \cite{mazur1970harmonic,suzuki1971}, who studied the magnetization temporal autocorrelation function.  The effect of  magnetic field on acoustic phonon modes have also been considered experimentally~\cite{sytcheva2010acoustic}.
 
 In the present work we consider a charged harmonic chain of $N$ oscillators in the presence of a uniform magnetic field and with  ends connected to  heat reservoirs  at different temperatures. The baths are modeled via white noise Langevin equations so that the steady state is Gaussian. We obtain the exact steady state current using the Langevin equations-NEGF formalism~\cite{dhar2006heat}.  In this formalism, the expression for the heat current, $J_N$, for a finite system size is obtained as an integral over heat transmitted at different frequencies. Therefore, one can then explore the  behaviour of the heat transmission coefficent, $\mathcal{T}_N(\omega)$ at different phonon frequencies $\omega$. In the absence of the magnetic field, heat transmission in our model takes place via independent transverse phonon modes. An interesting effect we observe is that the magnetic field couples the phonon modes, thus one finds that the NEGF heat transmission coefficient involves two Green's functions which separately contribute to the current. These two contributions can be interpreted physically as scattering processes between the two transverse plane wave modes which leads to change of polarization of the incoming plane wave.   We also derive explicit expressions, for the  effective transmission coefficients,  $\mathcal{T}_\infty$, and the integrated total current, $J_\infty$, in the  thermodynamic limit. We show that the magnetic field modifies the behaviour of the transmission at small frequencies,  giving  $\mathcal{T}_\infty \sim \omega^{1/2}$ and $\mathcal{T}_\infty \sim \omega^{3/2}$ for free and fixed boundaries respectively.  This is in contrast to the forms $\mathcal{T}_\infty \sim \omega^0,~\omega^2$ observed in the zero field case, for free and fixed boundary conditions respectively~\cite{roy2008heat}.

The paper is structured as follows: in Sec.~\ref{sec:model} we introduce the model and derive the heat current expression using the Langevin equations-NEGF formalism. The transmission coefficents involve particular Green's function elements and in Sec.~\ref{sec:green_prm}, we express these in terms of product of $2\cross 2$ matrices. In Sec.~\ref{sec:size_current} these are then  used to find the exact expressions for the current  in the thermodynamic limit for the case of uniform magnetic field. This is done for free as well as fixed boundary conditions. We conclude in Sec.~\ref{sec:concl}.

\section{The Model and derivation of heat current using NEGF}
\label{sec:model}
\subsection{The model}
We consider a  chain of $N$  harmonic oscillators each having two transverse degree's of freedom so that every oscillator is free to move in a plane perpendicular to the length of the chain. We choose the plane of motion to be the $x-y$ plane and  denote  the positions and momenta of the  $n^{th}$ oscillator by $(x_n,y_n)$ and $(p_{n}^x,p_{n}^y)$ respectively, with $n=1,2,\ldots,N$. The oscillators are assumed to have  masses, $m$, and each carry a positive charge $e$. We consider a site-dependent magnetic field $\vec B_n=B_n {\vec{\mathbf e}}_z$, perpendicular to the plane of motion, which can be obtained from a vector potential $\vec {\bf A}_n=(-B_n y_n,B_n x_n,0)$ at each lattice site. The Hamiltonian of the chain is given by:
\begin{align*}
H=\sum_{n=1}^N \frac{(p_{n}^x+e B_n y_n)^2  +(p_{n}^y-e B_n x_n)^2}{2m}&+ k \sum_{n=0}^N  \frac{(x_{n+1}-x_n)^2 + (y_{n+1}-y_n)^2}{2}
\end{align*} 
where $k$  denotes the inter particle spring constant. We will consider the two different boundary conditions: (i) fixed boundaries with $x_0=x_{N+1}=0$ and (ii) free boundaries with $x_0=x_1,~x_N=x_{N+1}$. In order to study heat current through this system, we consider the $1^{\rm st}$ and the $N^{\rm th}$ oscillators to be connected to heat reservoirs at temperatures $T_L$ and $T_R$ respectively. The heat reservoirs are modelled using dissipative and noise terms leading to the  following Langevin equations of motion:
\begin{align}	
m \ddot x_n&=k (x_{n+1}+x_{n-1}-c_n x_{n})+ e B_n \dot y_n+\eta_L^x(t)\delta_{n,1}+\eta_R^x(t)\delta_{n,N}-(\gamma \delta_{n,1}+\gamma\delta_{n,N})\dot x_n \label{xiddot},\\
m\ddot y_n&=k (y_{n+1}+y_{n-1}-c_n y_{n})- e B_n \dot x_n +\eta_L^y(t)\delta_{n,1}+\eta_R^y(t)\delta_{n,N}-(\gamma \delta_{n,1}+\gamma\delta_{n,N})\dot y_n \label{yiddot}.
\end{align}
for $n=1,2,\ldots,N$. Here $\eta_L (t) :=(\eta_L^x(t),\eta_L^y(t))$ and  $\eta_R (t):=(\eta_R^x(t),\eta_R^y(t))$ are Gaussian white noise terms acting on the  $1^{\rm st}$ and $N^{\rm th}$ oscillators respectively. These follow the regular white noise correlations, $\expval{\eta_{L/R}(t)\eta_{L/R}(t')}=2\gamma T_{L/R}\delta(t-t')$ (Boltzmann's constant is fixed to one to simplify), where $\gamma$   is the dissipation strength at the reservoirs.   The coefficients $c_n$ fix the boundary conditions of the problem. For fixed boundaries $c_n=2$ for all $n$, while for free boundary conditions $c_n=2-\delta_{n,1}-\delta_{n,N}$.

\subsection{Heat current using NEGF}
For heat current in the setup considered here, we need to obtain the steady state solution of the equations of motion given by Eq.~\eqref{xiddot}-\eqref{yiddot}. Denoting by ${\tilde u} (\omega)= \frac{1}{2\pi} \int_{-\infty}^\infty e^{-i \omega t} u(t) dt$ the Fourier transform of any function $u(t)$, we rewrite the Langevin equations in Fourier space as
 \begin{align}	
 (-m\omega^2+c_n k+i\gamma \omega\delta_{n,1}+i\gamma \omega\delta_{n,N})  \notag&\tilde x_n (\omega)-i \omega e B_n  \tilde y_n (\omega)\\&-k \tilde x_{n+1}(\omega)-k \tilde x_{n-1}(\omega)=\tilde \eta_L^x(\omega)\delta_{n,1}+\eta_R^x(\omega)\delta_{n,N} \label{qlex},\\
 (-m\omega^2+c_n k+i\gamma \omega\delta_{n,1}+i\gamma \omega\delta_{n,N}) \notag& \tilde y_n (\omega)+i \omega e B_n  \tilde x_n (\omega)\\&-k \tilde y_{n+1}(\omega)-k \tilde y_{n-1}(\omega)=\tilde \eta_L^y(\omega)\delta_{n,1}+\eta_R^y(\omega)\delta_{n,N}. \label{qley}
 \end{align}
The noise correlations in Fourier space now satisfy  $\expval{\eta_{L/R}(\omega)\eta_{L/R}(\omega')}=(\gamma T_{L/R}/\pi) \delta(\omega+\omega')$. 
Let us define the column vectors  
 \begin{equation*}
 \tilde X(\omega)=\begin{pmatrix}\tilde x_1(\omega)\\\tilde x_2(\omega)\\\vdots\\ \tilde{x}_{N-1}(\omega)\\\tilde x_N(\omega)\end{pmatrix} ~,~\tilde Y(\omega)=\begin{pmatrix}\tilde y_1(\omega)\\\tilde y_2(\omega)\\\vdots\\\tilde y_{N-1}(\omega)\\ \tilde y_{N}(\omega)\end{pmatrix}~,~\tilde\eta^{x/y}(\omega)=\begin{pmatrix}\tilde\eta_L^{x/y}(\omega)\\0\\\vdots\\ 0\\\tilde\eta_R^{x/y}(\omega)\end{pmatrix}
 \end{equation*}
 to write Eq.~\eqref{qlex} and Eq.~\eqref{qley} jointly in the block-matrix form
\begin{align}
\tilde{\mathcal{G}}^{-1}(\omega)
\begin{pmatrix}\tilde X(\omega)\\ \tilde Y(\omega)
\end{pmatrix}=\begin{pmatrix}
\tilde \eta^x(\omega)\\ \tilde \eta^y(\omega)
\end{pmatrix} \quad \text{with} \quad \tilde{\mathcal{G}}^{-1}(\omega)=\begin{pmatrix}
\Pi(\omega) && K(\omega)\\
-K(\omega)  && \Pi(\omega)
\end{pmatrix},
\label{bleq}
\end{align}
 where $\Pi(\omega)$ and $K(\omega)$ are square matrices with matrix elements given by
 \begin{align*}
 [\Pi(\omega)]_{n,\ell}&=(-m \omega^2+c_n k)\delta_{n,\ell}-k(\delta_{n,\ell+1}+\delta_{n,\ell-1})+i\gamma \omega\delta_{n,1}\delta_{\ell,1}+i\gamma \omega\delta_{n, N}\delta_{\ell, N}\ , \\
 [K (\omega)]_{n, \ell}&=-i e B_n \omega \delta_{n, \ell}.
 \end{align*}
 From Eq.~(\ref{bleq}), we can write the steady state solution directly by inverting the matrix $ \tilde{\mathcal{G}}^{-1}(\omega)$. Therefore we have
 \begin{align}
 \tilde x_n (\omega)&= \sum_\ell [G_1^+(\omega)]_{n, \ell} \ \tilde\eta_\ell^{x}(\omega)\; +\; \sum_\ell [G_2^+(\omega)]_{n, \ell} \ \tilde\eta_\ell^{y}(\omega),\label{xtildei}\\
 \tilde y_n (\omega)&= -\sum_\ell [G_2^+(\omega)]_{n,\ell} \ \tilde\eta_\ell^{x}(\omega)\; + \; \sum_\ell [G_1^+(\omega)]_{n, \ell} \ \tilde\eta_\ell^{y}(\omega),\label{ytildei}
 \end{align}
 where
 \begin{equation}
 G^+_1=\left[ \Pi +K \Pi^{-1} K\right]^{-1} ~~\text{and}~~G_2^+ =-G_1^+ K  \Pi^{-1}.
 \label{Gfeqns}
 \end{equation}
 These two last matrices form the $2\cross 2$ block structure of the matrix ${\tilde{\mathcal{G}}} (\omega)$ as
 \begin{equation*}
\tilde{\mathcal{G}}(\omega)=\begin{pmatrix}
 G_1^+(\omega) && G_2^+(\omega)\\
 -G_2^+(\omega)  && G_1^+(\omega)
 \end{pmatrix} .	
\end{equation*}
Defining the square matrices $G_{1/2}^- =[G_{1/2}^+]^\dagger$ and  $\Gamma_{n, \ell} (\omega)=[\Pi^\dagger(\omega)-\Pi(\omega)]_{n, \ell}=-2i\omega(\gamma\delta_{n,1}\delta_{\ell,1}+\gamma\delta_{n, N}\delta_{\ell,N})$, 
one  gets from Eq.~\eqref{Gfeqns} that $[G_1^-]^{-1} -[G_1^+]^{-1}=\Pi^\dagger-\Pi +K {\Pi^\dagger}^{-1}K-K \Pi^{-1}K  $. Multiplying this on the left by $G_1^{+}$ and on the right by $G_1^{-}$ we get, after some manipulations, the following relation: 
 \begin{equation}
 G^+_1(\omega)-G^-_1(\omega)=G_1^+\Gamma G_1^- + G_2^+ \Gamma G_2^- .
 \label{diffG}
 \end{equation}
This  will be useful later on to put the heat current in the  Landauer or NEGF form.
 
Having obtained the steady state solution, we now proceed to calculate the average heat current $J_N$ in the steady state.  We can compute the current at any point on the chain since the steady state value will be the same everywhere. 
Let us consider the current from the left reservoir into the system. This is given by taking the steady state average  $\langle \cdot\rangle_{\rm{ss}}$ of the dot product of the velocity of the first particle with the force on it from the left reservoir, thus
 \begin{equation*}
J_N =   -\gamma  \expval{\dot x_1^2+\dot y_1^2}_{\rm_{ss}} +\langle\eta_L^x(t) \dot  x_1\rangle_{\rm_{ss}}+\langle\eta_L^y(t) \dot  y_1\rangle_{\rm_{ss}}.
 \end{equation*} 
 The first term on  simplification gives
 \begin{align}
 -\gamma \expval{\dot x_1^2+\dot y_1^2}_{\rm_{ss}} =-\int_{-\infty}^{\infty} \frac{d\omega}{\pi}&\left\{ 2\gamma^2  T_L \omega^2 \left(\abs{[G_1^+(\omega)]_{1,1}}^2+\abs{[G_2^+(\omega)]_{1,1}}^2 \right) \right.\notag\\
 &\hspace{1cm}\left.+  2\gamma^2  T_R \omega^2 \left(\abs{[G_1^+(\omega)]_{1,N}}^2+\abs{[G_2^+(\omega)]_{1,N}}^2 \right)\right\}\label{curr1},
 \end{align}
  and the sum of the other two terms is given by
 \begin{align}
 &\langle\eta_L^x(t) \dot  x_1\rangle_{\rm_{ss}} +\langle\eta_L^y(t) \dot  y_1\rangle_{\rm_{ss}} =\int_{-\infty}^{\infty} \frac{d\omega}{\pi} \left\{ i\omega [G_1^+(\omega)]_{1,1} \gamma T_L  +	i\omega [G_1^+(\omega)]_{1,1} \gamma T_L \right\}\\
 &=2 T_L \int_{-\infty}^{\infty}  \frac{d\omega}{\pi} \omega^2 \left\{ \left(\abs{[G_1^+(\omega)]_{1,1}}^2+\abs{[G_2^+(\omega)]_{1,1}}^2\right)\gamma^2+(\abs{[G_1^+(\omega)]_{1,N}}^2+\abs{[G_2^+(\omega)]_{1,N}}^2)\gamma^2 \right\} 
 \label{curr2},
 \end{align}
where in the last step we used Eq.~\eqref{diffG}. Adding Eq.~\eqref{curr1} and Eq.~\eqref{curr2} we finally get
 \begin{align}
{J_N}&
=(T_L-T_R)\int_{-\infty}^{\infty} d\omega \mathcal{T}_N(\omega)
\label{eq:current007} 
 \end{align}
where ${\mathcal T}_N$ is the net transmission amplitude across the harmonic chain defined by
\begin{equation}
\mathcal{T}_N(\omega)= \frac{2 \gamma^2}{\pi}\, \omega^2\,\left[\abs{[G_1^+(\omega)]_{1,N}}^2+\abs{[G_2^+(\omega)]_{1,N}}^2\right]. \label{eq:TNfinal}
\end{equation}
Note that in the absence of the magnetic field, $B_n=0$, the Green's function, $G_2^+$, vanishes and we recover the current due to two uncoupled oscillator chains~\cite{roy2008heat}. The magnetic field couples the two transverse modes. In fact, we see from  Eq.~\eqref{xtildei} that $G_2^+(\omega)$ connects the $x$-displacements with the $y$-displacements, and hence,  it can be interpreted as the scattering matrix for a $x$-polarized incident plane wave to be scattered into a $y$-polarized wave. The term involving $G_1^+(\omega)$ is the normal transmission amplitude which is attributed to scattering of the incoming plane wave without  change of  polarization. The combination of the two terms leads to the rotation of the polarization of the incoming plane wave. 

It is interesting to note that the mathematical structure of the Green's functions obtained here is of  the same form as  that found for electron transport in superconducting wires~\cite{bhat2021,bhat2020transport}. Analogous to the scattering between transverse modes that we see here,  in superconducting wires, the superconducting order causes scattering between particle and hole electronic states.

\section{Green's function as product of matrices}
\label{sec:green_prm}
  In this section we rewrite the components of the two Green's functions, $[G_1^+(\omega)]_{1,N}$ and $[G_2^+(\omega)]_{1,N}$ defined by Eq.~\eqref{Gfeqns}, as a product of  matrices using a transfer matrix approach. This will give us explicit expressions for the two components enabling us to obtain analytic closed-form results in special cases. We start by rewriting the equations of motion in a way such that the $2N\cross 2N$ matrix $\tilde{\mathcal{G}}^{-1}(\omega)$ appearing in Eq.~\eqref{bleq} gets restructured  into a $2\cross 2$-block tri-diagonal matrix, $\mathcal{G}(\omega)$. To that end, we define for each $k=1, \ldots, N$ the column vectors
  \begin{align*}
  \tilde{R}_n (\omega)=\begin{pmatrix} \tilde x_n (\omega)\\  \tilde y_n (\omega) \end{pmatrix}, \quad  \tilde \eta_n (\omega)=\begin{pmatrix}
  \tilde \eta_n^x(\omega)\\\tilde \eta_n^y(\omega)
  \end{pmatrix} ,
  \end{align*}
 and notice that the Eq.~\eqref{qlex} and Eq.~\eqref{qley} can then be written as
\begin{equation}
\sum_{\ell=1}^N [\mathcal{G}^{-1}(\omega)]_{n, \ell}\tilde R_\ell (\omega)=\tilde \eta_n (\omega),
\label{Rjomega}
\end{equation}
 where $[\mathcal{G}^{-1}(\omega)]_{k, \ell}$ are $2\cross 2$ matrices defined via   $\mathcal{G}^{-1}(\omega)$ given by
\begin{equation*}
\mathcal{G}^{-1}(\omega)=\begin{pmatrix}
A_{1} (\omega)+i \gamma \omega I_2 & -k I_2&0  &0& \dots\dots& 0&0 \\
-k I_2 & A_{2} (\omega)  & -k I_2&0 & \dots\dots&0&0 \\
0 & -k I_2 & A_{3} (\omega) & -k I_2 &\dots \dots &0&0\\
\vdots&\vdots &\vdots & \ddots&&\vdots&\vdots \\
\vdots&\vdots&\vdots& & \ddots &\vdots&\vdots \\
\vdots&\vdots&\vdots&\dots& -k I_2&A_{N-1} (\omega) & -k I_2\\
0&0 & 0&\dots&\dots&-k I_2 &A_{N} (\omega)+i \gamma \omega I_2
\end{pmatrix}.
\end{equation*}
Here $I_n$ refers to a $n\cross n$ identity matrix, while $A_\ell (\omega)$ is the $2\cross 2$ matrix defined by
\begin{align*}
A_\ell:=A_{\ell} (\omega) =\begin{pmatrix}
-m\omega^2+c_\ell k && -i\omega eB_\ell \\
i\omega eB_\ell &&-m\omega^2+c_\ell k
\end{pmatrix}, \quad \ell=1, \ldots, N.
\end{align*}
Notice that Eq.~\eqref{Rjomega} gives $\tilde{R}_\ell (\omega)=\sum_{k} [\mathcal{G}(\omega)]_{\ell, k}\tilde{\eta}_k(\omega)$, so on comparison with the solution in Eq.~\eqref{xtildei} and Eq.~\eqref{ytildei} we can write the components of the $2\cross2$ matrix $[\mathcal{G}(\omega)]_{k, \ell}$ to be
\begin{equation}
[\mathcal{G}(\omega)]_{n, \ell}=\begin{pmatrix}[G_1^+(\omega)]_{n, \ell} && [G_2^+(\omega)]_{n, \ell}\\
-[G_2^+(\omega)]_{n, \ell} && [G_1^+(\omega)]_{n, \ell}
\label{mcG-G1G2}
\end{pmatrix}.
\end{equation}
Thus we now require the $2\cross2$ block  $[\mathcal{G}(\omega)]_{1,N}$ to calculate the components $[G_1^+(\omega)]_{1,N}$ and $[G_2^+(\omega)]_{1,N}$. Since the matrix $\mathcal{G}^{-1}(\omega)$ is tri-diagonal, $[\mathcal{G}_1^+(\omega)]_{1,N}$  can be expressed as products of matrices using a transfer matrix approach. This may be achieved by writing down the first column of equations from the  identity $\mathcal{G}(\omega)\mathcal{G}^{-1}(\omega)=I_{2N}$, 
\begin{align}
\mathcal{G}_{1,1}(A_{1}+ i {\gamma}\omega I_2 )-k \mathcal{G}_{1,2}&=I_2~,\label{Gstart}\\
 \mathcal{G}_{1,\ell-1}+  \mathcal{G}_{1,\ell+1} - k^{-1} \mathcal{G}_{1,\ell}A_{\ell} &=0~,~1<\ell<N\label{Gmid},\\
\mathcal{G}_{1,N}(A_{N}+ i \gamma \omega I_2 ) -k \mathcal{G}_{1,N-1}&=0\label{Gend}.
\end{align}
We now write these equations in the form
\begin{align}
\begin{pmatrix}
I_2 && \mathcal{G}_{11}
\end{pmatrix}&=\begin{pmatrix}
\mathcal{G}_{11} && \mathcal{G}_{12}
\end{pmatrix}\begin{pmatrix} A_{1}+i\gamma \omega I_2 &&I_2\\ -k I_2&&0\end{pmatrix},\label{Gstart1}\\
\begin{pmatrix}
\mathcal{G}_{1, \ell-1} && \mathcal{G}_{\ell,l}
\end{pmatrix}&=\begin{pmatrix}
\mathcal{G}_{1,\ell} && \mathcal{G}_{1,\ell+1}
\end{pmatrix}\begin{pmatrix} k^{-1} A_\ell && I_2 \\ - I_2&&0\end{pmatrix}~,~1<\ell<N\label{Gmid1}\\
\begin{pmatrix}
\mathcal{G}_{1,N-1} && \mathcal{G}_{1,N}
\end{pmatrix}&=\begin{pmatrix}
\mathcal{G}_{1,N} && 0
\end{pmatrix}\begin{pmatrix}k^{-1} (A_{N}+i\gamma \omega I_2) &&I_2\\ 0&&0\end{pmatrix}\label{Gend1}.
\end{align}
 We then  use Eq.~\eqref{Gmid1} in Eq.~\eqref{Gstart1} repeatedly  and finally use Eq.~\eqref{Gend1} to get,

\begin{align}
\begin{pmatrix}
I_2 & \mathcal{G}_{11}
\end{pmatrix}=\begin{pmatrix}
\mathcal{G}_{1N} & 0
\end{pmatrix}
\; \Omega_L \ \Pi_N \  \Omega_{R}
\label{gfprm}
\end{align}
with the the $4\cross 4$ matrices $\Omega_L, \Pi_N, \Omega_{R}$ defined by 
\begin{equation}
\begin{split}
\label{Omega}
&\Omega_L=\begin{pmatrix} I_2&&-i\tfrac{\gamma\omega}{k}  I_2\\0&&0 \end{pmatrix}, \quad \Omega_{R}=\begin{pmatrix} k I_2&&0\\i\gamma\omega I_2&&I_2\end{pmatrix},\\
&\Pi_N=\prod_{\ell=1}^{N}\Omega_\ell=\prod_{\ell=1}^{N}\begin{pmatrix} k^{-1} A_{\ell}&&I_2\\-I_2&&0\end{pmatrix}=\prod_{\ell=1}^{N}\begin{pmatrix} a_\ell I_2+b_\ell \sigma^y&&I_2\\-I_2&&0\end{pmatrix}
\end{split}
\end{equation}
where $\sigma^y=\left(\begin{smallmatrix}0&& -i\\ i&&0 \end{smallmatrix}\right)$ is the second Pauli's matrix and
\begin{equation}
\label{eq:aellb}
a_\ell:=a_\ell (\omega)=(-m\omega^2+c_\ell k)/k, \quad b_\ell:=b_\ell (\omega)=\omega e B_\ell /k.
\end{equation}

Thus we have expressed the required components of the Green's function as a product of $4\cross4$ matrices $\Omega_\ell$. This product can further be simplified by making a unitary transformation such that $\sigma^y =U^\dagger\sigma^zU$ with $\sigma^z=\left(\begin{smallmatrix}1&& 0\\ 0&&-1\end{smallmatrix}\right)$ the third Pauli's matrix, in order to diagonalise the matrix $a_\ell I_2+b_\ell\sigma^y$  . This makes the product to be
  \begin{align*}
  \Pi_N&=\prod_{\ell=1}^{N}\begin{pmatrix} a_\ell I_2+b_\ell \sigma^y&&I_2\\-I_2&&0\end{pmatrix} =\begin{pmatrix}U^\dagger &&0 \\ 0&& U^\dagger
  \end{pmatrix}\prod_{\ell=1}^{N}\begin{pmatrix} a_\ell I_2+b_\ell \sigma^z&&I_2\\-I_2&&0\end{pmatrix}\begin{pmatrix}U &&0 \\ 0&& U
  \end{pmatrix}.
  \end{align*}
  The  product in this equation is now composed of $2\cross 2$ diagonal blocks and therefore  we have that for any $1\le n\le N$, 
  \begin{align*}
  \prod_{\ell=1}^{n}\begin{pmatrix} a_\ell I_2+b_\ell \sigma^z&&I_2\\-I_2&&0\end{pmatrix}=
  \begin{pmatrix} 
  f_n^+ &&0 && g_{n}^+ &&0\\
  0 && f_n^- && 0 && g_{n}^-\\
  -f_{n-1}^+ &&0 && -g_{n-1}^+ &&0\\
  0 && -f_{n-1}^- && 0 && -g_{n-1}^-
  \end{pmatrix}
  \end{align*}
  where the numbers $f_n^\pm, g_n^\pm$, defined for $n=0,1,..,N$, follow the same second order recursive equation but with different initial conditions. More exactly we have that for 
  \begin{equation}
    \label{f_N}
  \begin{split}
 f_{n+1}^\pm&=(a_{n+1}\pm b_{n+1})f_n^\pm-f_{n-1}^\pm, \quad  f_0^\pm =1, \quad f_1^\pm =a_1\pm b_1,\\
 g_{n+1}^\pm&=(a_{n+1}\pm b_{n+1}) g_{n}^\pm-g_{n-1}^\pm, \quad g_0^\pm =0, \quad g_1^\pm =1.
\end{split}
\end{equation}

Therefore, the $4\cross4$ matrices in the product are effectively reduced to $2\cross 2$ matrices. The expressions for $f_n^\pm, g_n^\pm$ could be exactly found for the two boundary conditions. We do this in  the next section, for now we conclude this section by expressing required components of the Green's function using the variables $f_n^\pm, g_n^\pm$.  We use $U=\tfrac{1}{\sqrt{2}}\left(\begin{smallmatrix}i&&1\\-i&&1\end{smallmatrix}\right)$ to rewrite $\Pi_N$ as,

\begin{equation}
\label{Omega1}
\Pi_N=\frac{1}{2i}\setlength\arraycolsep{2pt}\begin{pmatrix} i(f_N^++f_N^-) &&( f_{N}^+-f_{N}^-) && i( g_N^++g_N^-) && ( g_{N}^+-g_{N}^-)\\
  -( f_{N}^+-f_{N}^-) &&i( f_N^++f_N^-) && -( g_{N}^+-g_{N}^-) &&i( g_N^++g_N^-)\\
  -i( f_{N-1}^++f_{N-1}^-) &&-( f_{N-1}^+-f_{N-1}^-) && -i( g_{N-1}^++g_{N-1}^-) &&-( g_{N-1}^+-g_{N-1}^-)\\
  ( f_{N-1}^+-f_{N-1}^-) &&-i( f_{N-1}^++f_{N-1}^-) && ( g_{N-1}^+-g_{N-1}^-) &&-i( g_{N-1}^++g_{N-1}^-)
  \end{pmatrix}.
  \end{equation}
We define  the matrices $P_N$ and $Q_N$ as follows,
\begin{equation}
    P_N=\frac{1}{2i}\setlength\arraycolsep{1.5pt}\begin{pmatrix}
  i(f_N^++f_N^-) && f_N^+-f_N^-\\-(f_N^+-f_N^-) && i(f_N^++f_N^-)\end{pmatrix} \quad Q_N=\frac{1}{2i}\setlength\arraycolsep{1.5pt}\begin{pmatrix}
  i(g_N^++g_N^-) && g_N^+-g_N^-\\-(g_N^+-g_N^-) && i(g_N^++g_N^-)\end{pmatrix}.
  \end{equation}
  Then 
  \begin{equation}
  \Pi_N=\begin{pmatrix}
  P_N && Q_N\\
  -P_{N-1}&& -Q_{N-1}
  \end{pmatrix}.
   \end{equation}
 Substituting $\Pi_N$ from Eq.~\eqref{Omega1} in  Eq.~\eqref{gfprm}, we can show that
  \begin{align}
\label{G1Neq}
  I_2&=\mathcal{G}_{1,N}\left(k P_N+i\gamma\omega(Q_N+P_{N-1})-\frac{\gamma^2\omega^2}{k}Q_{N-1}\right)=\mathcal{G}_{1, N}\begin{pmatrix}\frac{1}{2}(F_N^++F_N^-) && \frac{1}{2i}(F_N^+-F_N^-)\\-\frac{1}{2i}(F_N^+-F_N^-) && \frac{1}{2}(F_N^++F_N^-)\end{pmatrix}\\
  \mathcal{G}_{11}&=\mathcal{G}_{1, N}\left(Q_N+i\frac{\gamma \omega }{k}Q_{N-1}\right)\label{G11eq}
  \end{align}

 where 
 \begin{equation}
 \label{eq:FN}
  F_N^\pm :=F_N^\pm (\omega) = k \left( f_N^\pm+i\tfrac{\gamma}{k} \omega (g_{N}^\pm+f_{N-1}^\pm)-\tfrac{\gamma^2}{k^2}  \omega^2 g_{N-1}^\pm \right) .
\end{equation}
Using Eq.~\eqref{mcG-G1G2} and inverting Eq.~\eqref{G1Neq} gives us the required components of the Green's functions:
  \begin{equation*}
  [G_1^+]_{1,N}=\frac{1}{2}\left(\frac{1}{F_N^+}+\frac{1}{F_N^-}\right)~~\text{and}~~[G_2^+]_{1, N}=-\frac{1}{2i}\left(\frac{1}{F_N^+}-\frac{1}{F_N^-}\right).
  \end{equation*}
  These then give, using Eqs.~\eqref{eq:current007}-\eqref{eq:TNfinal}, the heat current to be
   \begin{equation}
   \label{eq:exprCurrent}
{J}_N =(T_L-T_R) \, \frac{ \gamma^2}{\pi}\int_{-\infty}^{\infty} d\omega \ \omega^2 \ \left[\frac{1}{\abs{F_N^+ (\omega)}^2}+\frac{1}{\abs{F_N^- (\omega) }^2}\right]. 
  \end{equation}
Thus we have now obtained a new expression for the net transmission amplitude ${\mathcal T}_N(\omega)$. In the next section we use this form and, for the case of a uniform magnetic field, derive analytical expressions for the current in the thermodynamic limit. Before that, we take a quick digression to discuss the temperature profile.
%

\subsection*{\it{Temperature Profile}}
\begin{figure}[t!]
	\centering
	\subfigure[$B=0$]{
		\includegraphics[width=7.5cm,height=6cm]{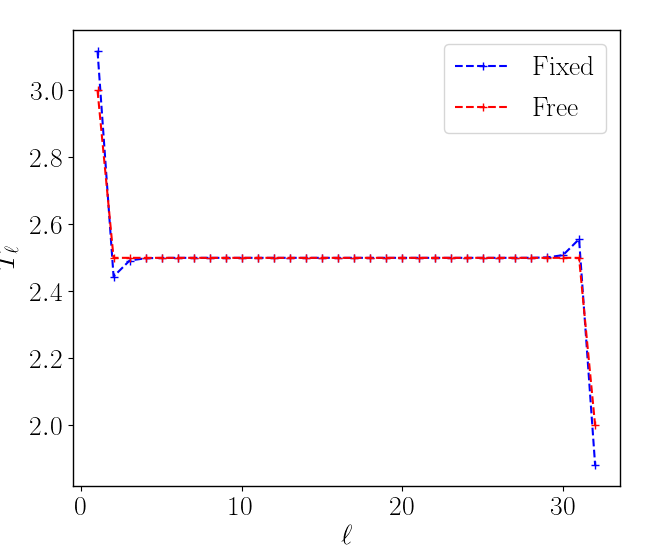}
	}
	\subfigure[$B=0.5$]{
		\includegraphics[width=7.5cm,height=6cm]{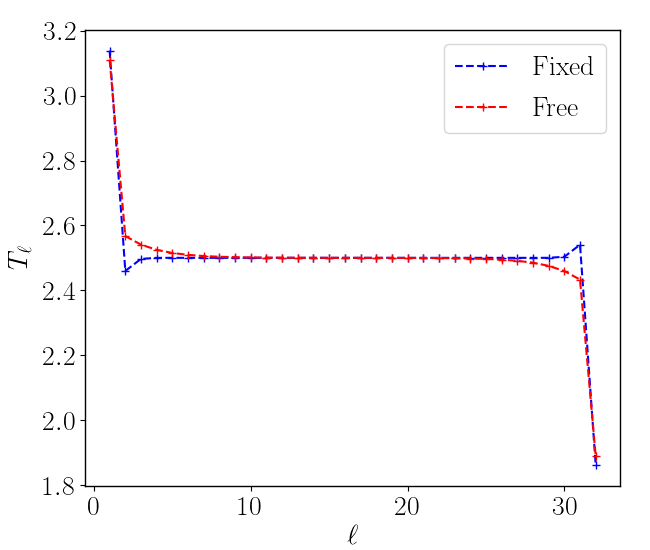}
	}	
	\caption{Temperature distribution for the ordered chain. Parameter values --- $e=m=k=1, N=32, T_L=3.5, T_R=1.5$.}\label{tempfig}
\end{figure}

We can also obtain the temperature profile of the chain, which is defined as $T_\ell=m\expval{\dot x_\ell^2(t)+\dot y_\ell^2(t)}$. Using the steady state expression for $x_\ell(t)$ and $y_\ell(t)$, we can show that this is given by,
\begin{equation}
T_\ell= T_L \Lambda_\ell+T_R(1-\Lambda_\ell);~~\Lambda_\ell=\int_{-\infty}^{\infty}\frac{d\omega}{\pi}m\gamma\omega^2\left[[|G_1^+(\omega)]_{1 \ell }|^2+|G_2^+(\omega)]_{1 \ell }|^2\right]
\end{equation}   
Using Eq.~(\ref{Gstart1}), Eq.~\eqref{Gmid1} and Eq.~\eqref{Gend1}, we could obtain the matrix block $\mathcal{G}_{1\ell}$, containing the required components for  the temperature profile, to be, 
\begin{align}
\begin{pmatrix}
I_2 & \mathcal{G}_{11}
\end{pmatrix}=\begin{pmatrix}
\mathcal{G}_{1\ell} & \mathcal{G}_{1,\ell+1}
\end{pmatrix}
\;  \Pi_\ell  \Omega_{R}=\begin{pmatrix}
\mathcal{G}_{1\ell} & \mathcal{G}_{1,\ell+1}
\end{pmatrix}\begin{pmatrix}
kP_\ell+i\gamma\omega Q_\ell  && Q_\ell\\
-kP_{\ell-1}-i\gamma\omega Q_{\ell+1} && -Q_{\ell-1} 
\end{pmatrix}
\end{align}  
where $\Pi_\ell$ is defined by Eq.~\eqref{Omega1} with $n$ replaced by $\ell$. Using this equation we can write linear equations for $\mathcal{G}_{1\ell}$ and $\mathcal{G}_{1,\ell+1}$ in the block form as,
\begin{align}
	I_2-i\gamma\omega\mathcal{G}_{11}=\mathcal{G}_{1\ell}kP_\ell-\mathcal{G}_{1,\ell+1}kP_{\ell-1}\ ,\\
	\mathcal{G}_{11}=\mathcal{G}_{1\ell}Q_\ell-\mathcal{G}_{1,\ell+1}Q_{\ell-1}\ .
\end{align}
$\mathcal{G}_{11}$ is obtained via Eq.~\eqref{G11eq}. This set of equation seems complicated to simplify further but when evaluated numerically, for an ordered chain, we obtain the results given in Fig~(\ref{tempfig}). The temperature in the bulk is the same as for zero magnetic field case, $(T_L+T_R)/2$ and the magnetic field mostly effects the profile near the ends of the chain.

\section{Expressions for the Current in the Thermodynamic limit}
\label{sec:size_current}
We consider a uniform chain ($B_n=B$ for all $n=1,2\ldots,N$) and derive the expressions for the current in the thermodynamic limit, 
\begin{equation*}
{J}_{\infty}= \lim_{N \to \infty} {J}_N,
\end{equation*}
for the cases of fixed boundary conditions and free boundary conditions. 

\begin{figure}[t!]
	\centering
	\subfigure[$B=1$]{
		\includegraphics[width=7.5cm,height=7.5cm]{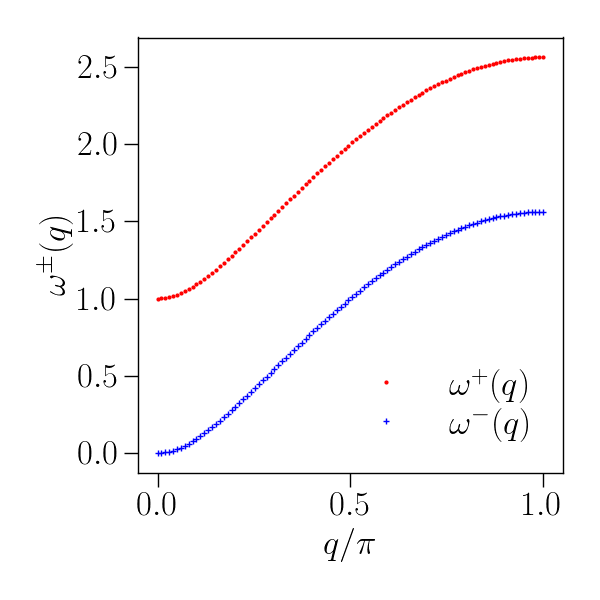}
	}
	\subfigure[$B=2$]{
		\includegraphics[width=7.5cm,height=7.5cm]{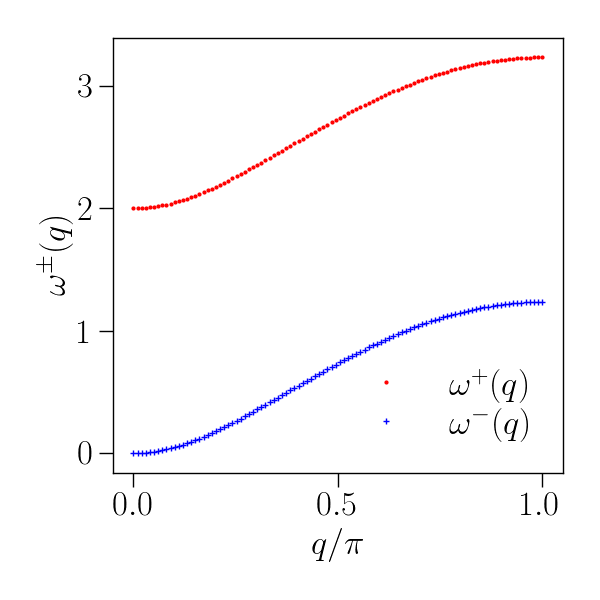}
	}	
\caption{Spectrum of the chain in the bulk. Parameter values --- $e=m=k=1$.}\label{spec}
\end{figure}

For the infinite system, the phonon spectrum consists of two bands $\{ \omega^\pm (q) \; ; \; q \in (0,\pi)\}$ where $2m\omega^\pm(q)=\pm eB+ \sqrt{(eB)^2+8mk(1-\cos q)}$. The bands are gapped for $eB>\sqrt{2mk}$. In Fig.~(\ref{spec}a) and Fig.~(\ref{spec}b) we show the spectrum for $eB<\sqrt{2mk}$ and $eB>\sqrt{2mk}$ respectively. In the former, the bands overlap while in the latter they are gapped.
We expect the transmission to be zero outside the bandwidth of the two bands which becomes explicit in the thermodynamic limit. We will show that in the thermodynamic limit, the current expression in Eq.~\eqref{eq:exprCurrent} is the sum of two terms due to the two frequency bands of the system. Also, in the small $\omega$ limit  we will find that $\mathcal{T}_\infty(\omega)=\lim_{N\rightarrow\infty}\mathcal{T}_N(\omega)$ is equivalent to $\omega^{3/2}$ and $\omega^{1/2}$ for free and open boundary conditions respectively. We consider the two boundary conditions separately and set $k=e=1$ without loosing generality.

 \subsubsection*{Fixed boundary conditions:} We have then $c_\ell=2$ for all $\ell=1, \ldots, N$. The expressions for $f_\ell^\pm$ for free boundary conditions can be found exactly from Eq.~\eqref{f_N}. Recalling Eq.~\eqref{eq:aellb} let us denote $q^\pm:=q^{\pm} (\omega) \in \mathbb C$ such that
 \begin{equation}
 \label{eq:defq}
2 \cos (q^\pm (\omega) ) = a_\ell (\omega) \pm b_\ell(\omega) =2- {m}\omega^2 \pm {B} \omega. 
\end{equation}
We obtain 
\begin{align}
\label{eq:fl}
 	f_\ell^\pm=\frac{\sin[q_\pm (\ell+1)]}{\sin q_\pm} \quad \text{and}\quad g^\pm_\ell = f^\pm_{\ell -1} \quad \text{(for $\ell\ge 1$)}. 
\end{align}

Recall Eq.~\eqref{eq:exprCurrent}. If $\omega$ is not in the frequencies band defined by $\omega^{+}$ (resp. $\omega^-$) then $q^{+} (\omega)$ (resp. $q^- (\omega)$) has a non-vanishing imaginary part and $F_N^+ (\omega)$ (resp. $F_N^- (\omega)$) becomes exponentially large in $N$, so that these $\omega$'s will not contribute in the thermodynamic limit to the heat current.

We therefore obtain in the thermodynamic limit,
 \begin{align*}
{J}_\infty&= (T_L -T_R) \ \frac{\gamma^2}{\pi} \lim\limits_{N\rightarrow\infty}\int_{-\infty}^{\infty}d\omega\  \omega^2 \  \left[\frac{1}{\abs{F_N^+ (\omega)}^2}+\frac{1}{\abs{F_N^- (\omega)}^2}\right]\\
&=(T_L -T_R) \ \frac{2\gamma^2}{\pi} \lim\limits_{N\rightarrow\infty}\left\{ \int_{0}^{\pi} d\omega^+ (q)    \frac{[\omega^+ (q)]^2}{\abs{F_N^+ (\omega^+ (q) )}^2} \ + \ \int_{0}^{\pi} d\omega^- (q)    \frac{[\omega^- (q)]^2}{\abs{F_N^- (\omega^- (q) )}^2}   \right\}
\end{align*}
where in the second equality, the $2$ comes from the fact that by symmetry we restricted us to positive frequencies.

To obtain the limits, we follow the steps given in \cite{roy2008heat}. By using Eq.~\eqref{eq:FN} and Eq.~\eqref{eq:fl} we can express $F_N^\pm (\omega^{\pm} (q))$ as
\begin{equation}
\label{eq:F_N0}
\begin{split}
	&F_N^\pm (\omega^{\pm} (q)) =\frac{1}{\sin (q)}[\alpha^{\pm}(q)\sin(q N)+ \beta^{\pm}(q) \cos(q N)],\\
	&\alpha^\pm(q)=(1-\gamma^2 [\omega^\pm (q)]^2) \cos(q)+2i\gamma \omega^\pm (q), \quad \beta^\pm(q)=(1+\gamma^2[\omega^\pm (q)]^ 2)\sin (q).
	\end{split}
\end{equation} 
We have then to study the limit as $N\to \infty$ of
\begin{equation*}
\begin{split}
&\int_0^\pi dq\ \cfrac{\partial \omega^\pm}{\partial q} (q) \cfrac{\sin^2 (q) [\omega^\pm (q)]^2}{\vert \alpha^{\pm}(q)\sin(q N)+ \beta^{\pm}(q) \cos(q N)\vert^2}=\int_0^\pi dq\ \cfrac{H^\pm (q)}{1 + R^\pm (q) \sin (2qN + \varphi^\pm (q)) }
\end{split}
\end{equation*}
where
\begin{equation*}
H^{\pm} (q) = \cfrac{2}{\vert \alpha^\pm (q)\vert^2 +\vert \beta^\pm (q)\vert^2}  \cfrac{\partial \omega^\pm}{\partial q} (q)  \ {\sin^2 (q) [\omega^\pm (q)]^2}
\end{equation*}
and 
\begin{equation*}
R^\pm (q) \cos (\varphi^\pm (q)) =\cfrac{2 \Re (\alpha^\pm (q) \overline{{\beta}^\pm} (q))}{\vert \alpha^\pm (q)\vert^2 +\vert \beta^\pm (q)\vert^2} \  , \quad R^\pm (q) \cos (\varphi^\pm (q)) =\cfrac{\vert \alpha^\pm (q)\vert^2 -\vert \beta^\pm (q)\vert^2}{\vert \alpha^\pm (q)\vert^2 +\vert \beta^\pm (q)\vert^2}.
\end{equation*}
$\overline{{\beta}^\pm}(q)$ is the complex conjugate of $\beta^\pm(q)$. It can be shown~\cite{roy2008heat,casher1971heat,kannan2012} that 
\begin{equation*}
\lim_{N\to \infty} \int_0^\pi dq\ \cfrac{H^\pm (q)}{1 + R^\pm (q) \sin (2qN + \varphi^\pm (q)) } =\int_0^\pi dq \ \cfrac{H^\pm (q)}{\sqrt{1-[R^\pm (q)]^2} }.
\end{equation*}
Since we have that 
$$\frac{1}{\sqrt{1-[R^{\pm}]^2}}=\frac{\vert \alpha^\pm\vert^2 +\vert \beta^\pm \vert^2}{ 2 \vert \Im (\alpha^\pm  {\overline{\beta^\pm}}) \vert}$$ 
and
\begin{equation*}
\Im \left(\alpha^\pm (q)  {\overline{\beta^\pm}} (q) \right) = 2\gamma \omega^\pm (q) \left( 1+\gamma^2 [\omega^\pm (q)]^2\right) \sin(q) \ge 0 \quad  \text{ for $q\in (0,\pi)$ },
\end{equation*}
we get that
\begin{equation}
\label{eq:fund1}
\begin{split}
&\lim_{N\to \infty} \int_0^\pi dq\ \cfrac{\partial \omega^\pm}{\partial q} (q) \cfrac{\sin^2 (q) [\omega^\pm (q)]^2}{\vert \alpha^{\pm}(q)\sin(q N)+ \beta^{\pm}(q) \cos(q N)\vert^2}\\
&=\cfrac{1}{2\gamma} \int_0^\pi dq\ \cfrac{\partial \omega^\pm}{\partial q} (q) \cfrac{ \omega^\pm (q) \sin(q)}{1+\gamma^2 [\omega^\pm (q)]^2} = \cfrac{1}{2\gamma} \int_{\omega^{\pm} (0)}^{\omega^\pm (\pi)} d\omega \  \cfrac{ \omega \sin(q^{\pm} (\omega) )}{1+\gamma^2 \omega^2}. 
\end{split}
\end{equation}
We conclude that
\begin{equation}
\label{currth}
\begin{split}
&{J}_\infty 
=(T_L -T_R) \  \cfrac{\gamma}{\pi} \left\{  \int_{\omega^{+} (0)}^{\omega^+ (\pi)} d\omega \  \cfrac{ \omega \sin(q^+ (\omega) )}{1+\gamma^2 \omega^2}   \; +\;  \int_{\omega^{-} (0)}^{\omega^- (\pi)} d\omega \  \cfrac{ \omega \sin(q^{-} (\omega) )}{1+\gamma^2 \omega^2}\right\}. 
\end{split}
\end{equation}
The two integrals run over the two bands of the spectrum: $[\omega^- (0), \omega^- (\pi)]=[0, (-B+\sqrt{B^2+16m})/2m]$ and $[\omega^+ (0), \omega^+ (\pi)]=[B/2m , (B+\sqrt{B^2+16m})/2m]$. We see that in the thermodynamic limit the transmission is exactly zero at energy values outside the two bands of the spectrum and also the current is explicitly expressed as sum of two terms coming from the two bands. For small $\omega$ behaviour of the transmission, $\mathcal{T}_\infty(\omega)$, we take the contribution due to the lower band (depending on the sign of $eB$ we have $\omega^+ (0)=0$ or $\omega^- (0)=0$). It is straightforward to see from Eq.~\eqref{currth} that $\mathcal{T}_\infty(\omega)\sim \omega^{3/2}$ for $B\neq 0$ while for $B=0$, $\mathcal{T}_\infty(\omega)\sim \omega^{2}$.

\begin{figure}[t!]
	\centering
	\subfigure[Fixed BC]{
		\includegraphics[width=7.5cm,height=7.5cm]{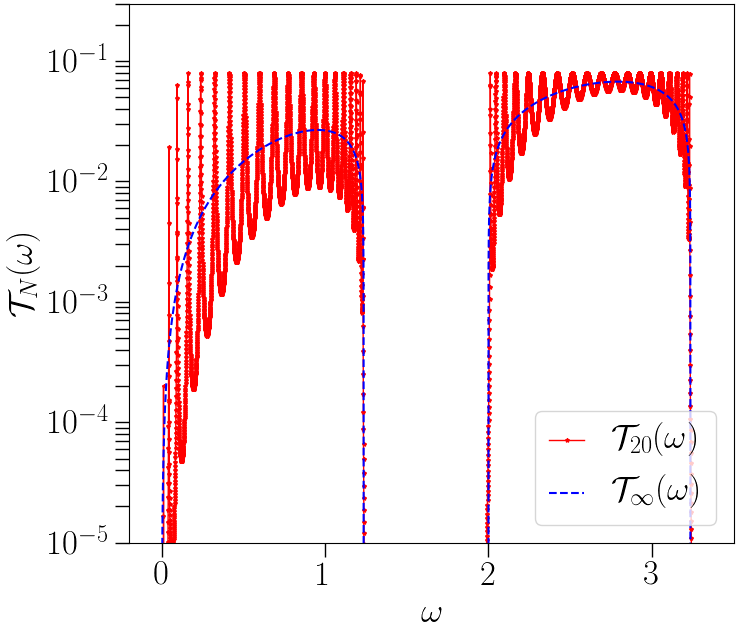}
	}	
	\subfigure[Free BC]{
		\includegraphics[width=7.5cm,height=7.5cm]{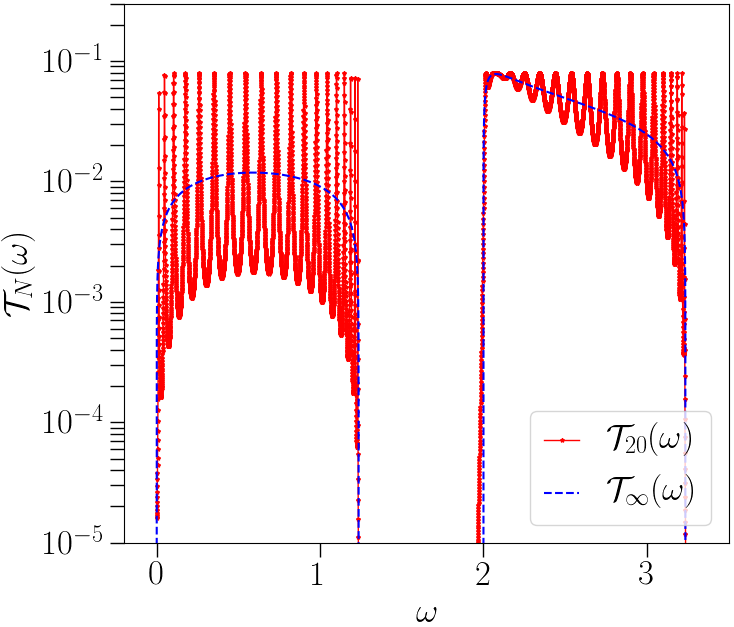}
	}
	\caption{Comparison of the transmission ${\mathcal T}_N (\omega)$ for fixed and free boundary for $N=20$ with $\mathcal{T}_\infty(\omega)$. Parameter values --- $m=k=e=1$, $\gamma=0.2$ and $B=2$.}\label{tau_plts1}
\end{figure}
\begin{center}
\begin{table}
\label{table}
\begin{center}
\begin{tabular}{|c|c|c|c|c|}
	\cline{1-5}
		& \multicolumn{2}{c|}{${J}_{N}$} & \multicolumn{2}{c|}{${J}_{\infty}$} \\ \cline{2-5} 
		& Fixed BC      & Free BC      & Fixed BC          & Free BC         \\ \hline\hline
		\multicolumn{1}{|c|}{$N=10$, $B=1$} &  0.179             &  0.160           &       0.179            &  0.158  \\ \hline
		\multicolumn{1}{|c|}{$N=20$, $B=1$} &  0.179           &   0.158           &       0.179            &    0.158   \\ \hline\hline
		\multicolumn{1}{|c|}{$N=10$, $B=2$} &                     0.163       &        0.121 &0.163            &        0.131 \\ \hline
		\multicolumn{1}{|c|}{$N=20$, $B=2$}    &             0.163     &          0.131 &     0.163       &        0.131  \\ \hline
\end{tabular}
\end{center}
\caption{Comparison of numerical values of the current for finite $N$ and the value of the current in the thermodynamic limit for $\gamma=0.2$, $m=k=e=1$, $T_L=1,T_R=0$.}
\end{table}
\end{center}
\begin{figure}[t!]
	\centering
	\subfigure{
		\includegraphics[width=9cm,height=7.5cm]{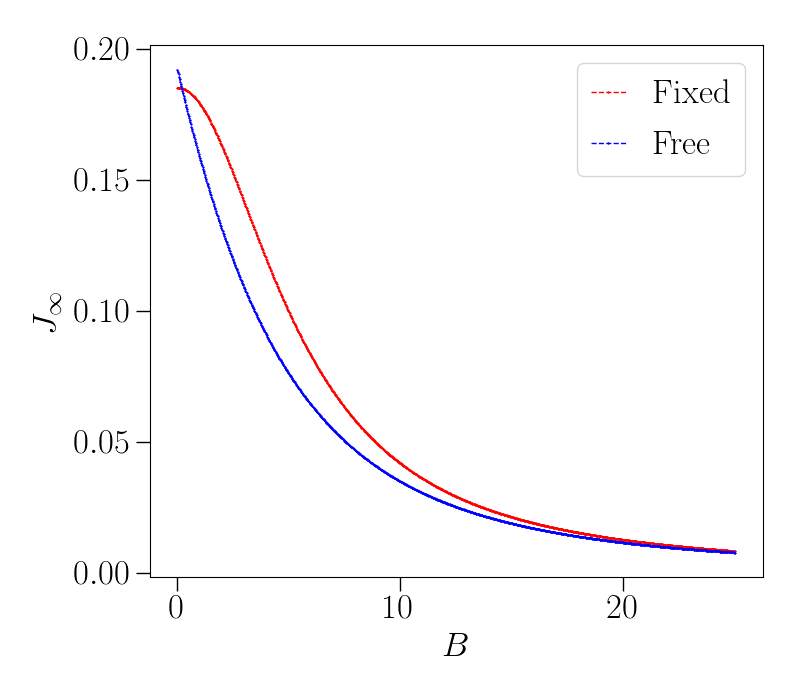}
	}
	\caption{Variation of the current with the magnetic field. Parameter values --- $e=m=k=1,\gamma=0.2$, $T_L=1,T_R=0$.}\label{curmag}
\end{figure}

\subsubsection*{Free boundary conditions:} For free boundary conditions we have $c_\ell=2-\delta_{\ell,1}-\delta_{\ell,N}$. Recalling Eq.~\eqref{eq:aellb} and the definition of $q^\pm:=q^{\pm} (\omega) \in \mathbb C$ the numbers $f_\ell^\pm, g_\ell^\pm$ can once again be obtained with from Eq.~\eqref{f_N}. We have
\begin{align*}
f_N^{\pm}&=2\frac{(\cos (q_\pm) -1)}{\sin (q_\pm) }\sin(q_\pm N),\quad g_{N-1}^{\pm}=\frac{\sin(q_\pm(N-1))}{\sin (q_\pm)},\\
&\text{and}\quad g_N^{\pm}=f_{N-1}^{\pm}=\frac{1}{\sin (q_\pm)}(\sin (q_\pm N)-\sin (q_\pm (N-1))),
\end{align*}
where $q^\pm$ is defined in Eq.~\eqref{eq:defq}. Using these we can express $F_N^\pm$ defined by Eq.~\eqref{eq:FN} as
\begin{equation*}
F_N^\pm (\omega^\pm (q))=\frac{1}{\sin (q) }[\alpha^\pm (q)\sin(q N)+ \beta^\pm (q) \cos(q N)]
\end{equation*} 
where
\begin{align*}
	\alpha^\pm (q)&=2(\cos (q) -1)-\gamma^2[\omega^{\pm} (q)]^2\cos (q)+2i\gamma\omega^\pm (q) (1- \cos (q) ),\\
	\beta^\pm (q)&=\gamma^2 [\omega^\pm (q)]^2 \sin (q) + 2i\gamma \omega^\pm (q) \sin (q) .
\end{align*}
It has the same form as $F^\pm_N$ appearing in Eq.~\eqref{eq:F_N0} but with different expressions for $\alpha^\pm$ and $\beta^\pm$. Hence using the same method, and noticing that 
\begin{equation*}
\Im \left( \alpha^\pm (q) {\overline{\beta^\pm}} (q) \right) = 2 \gamma [\omega^{\pm}(q) ]^2 \sin (q) \left[ \mp {B} +\left(\gamma^2 + m \right) \omega^\pm (q) \right] 
\end{equation*}
we deduce that 
\begin{equation}
\label{currthf}
{J}_\infty = (T_L- T_R) \ \cfrac{\gamma}{\pi} \left\{  \int_{\omega^{+} (0)}^{\omega^+ (\pi)} d\omega \  \cfrac{\sin(q^+ (\omega) )}{- {B} +\left(\gamma^2 +{m} \right) \omega}   \; +\;  \int_{\omega^{-} (0)}^{\omega^- (\pi)} d\omega \  \cfrac{ \sin(q^{-} (\omega) )}{ {B} +\left(\gamma^2 +{m} \right) \omega}\right\}. 
\end{equation}
As in the case of fixed boundary condition, we have expressed the current as the sum of two integrals running over the two bands of the spectrum. However, from this expression, for small $\omega$ behaviour of $\mathcal{T}_\infty(\omega)$, the lower band gives $\mathcal{T}_\infty(\omega) \sim \omega^{1/2}$ and $\sim \omega^0$  for $B\neq 0$ and $B=0$ respectively.

In Fig.~(\ref{tau_plts1}a) and Fig.~(\ref{tau_plts1}b), we show a comparison between $\mathcal{T}_\infty(\omega)$ derived for the two boundary conditions with the respect transmission obtained numerically for $N=20$. It can be seen that the transmission in the thermodynamic limit looks exactly like the envelope covered by the transmission for finite $N$. Table 1 shows the comparison of the numerically obtained current for $N=10, 20$ and $B=1,2$ with the value of the current calculated from the Eq.~(\ref{currth}) and Eq.~(\ref{currthf}) for the two boundary conditions respectively. These show a good agreement. We also show in Fig.~(\ref{curmag}) the variation of the current in thermodynamic limit $J_\infty$ with respect to the magnetic field and we find that it decreases monotonically to $0$ with the magnetic field $B$, as $1/B^2$ for large $B$, independently of the boundary conditions. We can also check easily that the limit $B\to 0$ and $N\to \infty$ commute, i.e. the limit of $J_\infty$ as $B\to 0$ is equal to the normalised current of the ordered harmonic chain without magnetic field considered in \cite{rieder1967properties,nakazawa1968energy,nakazawa1970lattice,roy2008heat}, for free and fixed boundary conditions.

\section{Conclusion and perspectives}
\label{sec:concl}
In conclusion we studied heat transport in an ordered harmonic chain in the presence of a uniform magnetic field. Using non-equilibrium Green's function formalism we found that the heat current has contribution from two different terms involving two different Green's functions $G_1^+(\omega)$ and $G_2^+(\omega)$. These can be interpreted physically as the transmission amplitude of a transverse plane wave  being scattered without or with the $\pi/2$ rotation of its polarization  respectively. This happens due to the fact that the magnetic field couples the $x$ and $y$ coordinates of the oscillators. We expressed the required components of the Green's functions  as a product of $2\cross2$ transfer matrices in which form one sees explicitly the contribution to the current from the two phonon bands. In the thermodynamic limit, the currents become $N$-independent and we  obtained analytic expressions for the current for free and fixed boundary conditions. These expressions show  that at small $\omega$ and $B\neq 0$, the transmission, $\mathcal{T}_\infty(\omega)\sim\omega^{3/2}$ for fixed boundary and  $\mathcal{T}_\infty(\omega)\sim\omega^{1/2}$ for free boundary. In the companion paper \cite{Cane2021} we extend these results to the case where the external magnetic field is random.\\

In \cite{GiardinKurchan05} the authors introduce a family of heat conduction models in the presence of a uniform or a non-uniform magnetic field, conserving energy and possibly momentum, and argue that in all cases Fourier's law is satisfied. In particular, they consider models with impulsive magnetic fields which act periodically in time like kicks and relate then their models to systems of discrete time coupled chaotic maps. In the high temperature limit the discrete time models are reasonably well approximated by a purely stochastic model satisfying Fourier's law. It seems that a similar treatment can be performed for our system resulting in a harmonic chain with a stochastic noise, very similar to the model introduced in \cite{BernardinOlla05}. It would be of interest to investigate these last questions in further details.

\section*{Acknowledgements}

A.D. and J.M.B. acknowledge support of the Department of Atomic Energy, Government of India, under Project No. RTI4001. The work of C.B. and G.C. has been supported by the projects LSD ANR-15-CE40-0020-01 of the French National Research Agency (ANR), by the European Research Council (ERC) under  the European Union's Horizon 2020 research and innovative programme (grant agreement No 715734) and the French-Indian UCA project `Large deviations and scaling limits theory for non-equilibrium systems'.

\bibliographystyle{halpha}
\bibliography{biblio}

\end{document}